\documentclass[smallextended]{svjour3}       
\smartqed  
\usepackage{graphicx}
\usepackage{layouts} 
\usepackage{amssymb} 
\usepackage{upgreek}
\begin{document}

\title{Surface Waves on Superfluid $^3$He and $^4$He%
}


\author{M.S.~Manninen         \and
        J.~Rysti              \and
        I.A.~Todoshchenko     \and
        J.T.~Tuoriniemi 
}


\institute{M.S.~Manninen \and J.~Rysti \and I.A.~Todoshchenko \and J.T.~Tuoriniemi \at
              Low Temperature Laboratory, Department of Applied Physics, Aalto University, FI-00076 AALTO, FINLAND \\
              \email{msmannin@boojum.hut.fi}           
}

\date{Received: \today / Accepted: date}

\maketitle

\begin{abstract} 

Surface waves on both superfluid $^3$He and $^4$He were examined with the premise, 
that these inviscid media would represent ideal realizations for this fluid dynamics problem.
The work on $^3$He is one of the first of its kind, but on $^4$He it was possible to produce much more complete set of data 
for meaningful comparison with theoretical models. Most measurements were performed at the zero temperature limit, 
meaning $T<$~100~mK for $^4$He and $T\sim$~100~$\mu$K for $^3$He. 
Dozens of surface wave resonances, including up to 11 overtones, 
were observed and monitored as the liquid depth in the cell was varied. 
Despite of the wealth of data, perfect agreement with the constructed theoretical models could not be achieved. 

%
\end{abstract}

\section{Introduction}
\label{sec:intro}

Fluid dynamics is an important branch of physics. However, accurate modeling of fluid flow in general geometries is cumbersome, and proper account for the effects of viscosity often leads to serious difficulties. Superfluids, on the other hand, offer the possibility to study quite common flow problems without the complications due to viscosity and, thus, they represent the closest possible realization of the ideal fluid often hypothesized in theoretical models. Also, the extraordinary wetting properties of the superfluids lead to well defined conditions in cases, where the free fluid surface comes to contact with solid walls. For these reasons, it was thought that the surface waves on the free surface of superfluids would obey a simple theoretical description nice and easy. In practice, however, it turned out not so straightforward to obtain good agreement with the experiment and the theory. This paper summarizes our understanding of the problem at present.

Surface waves on free fluid surfaces are perhaps the most easily perceived representation of wave phenomena in our common environment. The standing wave modes in restricted geometries can easily be demonstrated and analytical solutions to these in some idealized cases have been known for long. The resonance frequencies of the surface modes are influenced by kinetic, potential, and surface energies, and the problem in question can be viewed from several perspectives. Both bulk fluid and surface properties are connected together and they need to be incorporated to any proper model. It must be acknowledged that the free fluid surface in a bound geometry is not perfectly flat, but the menisci caused by the surface tension have an important contribution to the energy balance. Under general circumstances the vapor phase above the fluid surface would have a minor contribution to the problem but in the case of helium fluids at very low temperatures the gas phase becomes extremely rare and can be ignored completely.

Even if we are dealing with superfluids here and the fluid itself can flow without friction at low velocities, the surface waves still experience damping at any finite temperature. This is due to interaction of the moving interface with thermal quasiparticles in the fluid, and can be measured by observing the frequency width of the surface wave resonances. At sufficiently low temperatures the quasiparticles become ballistic, they scatter from the surfaces only, and their effect can be formulated easily. Interestingly, the quasiparticles have quite different character in bosonic $^4$He and in fermionic $^3$He superfluids, which lead to entirely different temperature dependencies. This feature has been presented in more detail in an earlier paper~\cite{Manninen14}, but will be reviewed shortly here again.

We studied standing waves on free fluid surface both in superfluid $^3$He and $^4$He. 
Surface waves in superfluid $^3$He have been studied only very recently~\cite{Eltsov13arxiv}
thus our work in $^3$He pushes to new frontiers with this rather exotic substance 
and the required microkelvin range temperatures.
However, our data on $^4$He is much more systematic and complete, 
and thus better suited for meaningful comparison with theoretical models. 
Nevertheless, interesting comparison can be made between the results on those two fluids investigated in the exact same experimental cell.

In this paper our focus is mainly on the resonance frequencies and energy balance in the zero temperature limit, though we give some remarks on the temperature dependencies of the characteristic quantities, and shortly discuss the surface waves above the superfluid transition temperature $T_c$. We were able to distinguish and identify several resonance frequency modes with varying depth of liquid helium in the system. The resonance frequencies were determined with great precision but the nontrivial geometry of our cell prevented us from deriving a conclusive value for surface tension in the zero temperature limit. Selected frequency bands could be fitted with credible values from the literature but the best overall correspondence across the whole data set led to somewhat contradicting values for some system parameters.

\section{Experimental setup}\label{sec:ExperimentalSetup}
\begin{figure}
\includegraphics{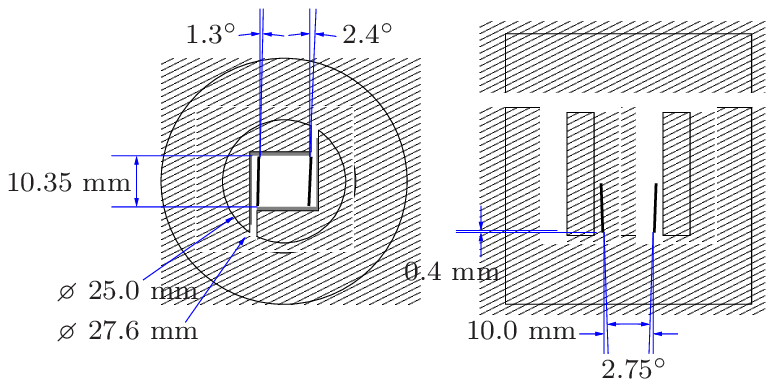}%
~~\includegraphics[width=0.3\textwidth]{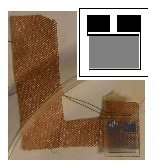}%
\caption{(Color online) Geometry of the experimental cell in top view (left) and side view (center). %
The annealed copper cell walls (dashed) were coated with about 0.5~mm thick layer of sintered silver powder heat exchanger (gray). %
The two interdigital capacitors (IDC) were located at the opposite vertical walls of the cuboid volume.
The full sensitive range of the IDCs is 4.6~mm in vertical direction and
it starts at about 1.4~mm from the porous sintered bottom
(or 2.0~mm from the impermeable copper bottom, where the thickness of the sintered layer is about 0.6~mm). %
The two capacitors were slightly tilted, as indicated by the measured angles. %
The IDCs were thermalized to the cell wall with copper mesh (right)
}
\label{fig:cell}       
\end{figure}

Oscillations of the free surface of liquid helium were studied in an experimental cell illustrated in Fig.~\ref{fig:cell}.
Heat was transferred from the helium sample via sintered silver powder heat exchangers
to the copper nuclear cooling stage of a microkelvin refrigerator~\cite{Yao00YKI}.
Temperature was measured at different stages by a germanium resistor, a SQUID-based noise thermometer, a platinum NMR thermometer, or deduced from the demagnetization field at the lowest temperatures.

The fluid surface level was sensed with an interdigital capacitor (IDC) on a vertical wall
of the cuboid volume with time resolution of about 10~ms, allowing detection of resonances below about 100~Hz frequency.
The detection signal amplitude had to be kept low for two reasons: first because large readout signals generated heat, which was detrimental for the experiment, in particular in the case of superfluid $^3$He, but also because large electric field/heated capacitor surface attracted superfluid on the surface and altered the conditions of the measurement. Level sensitivity of about 0.1~$\mu$m was achieved without notable undesired effects.
Another capacitor was located on the opposite wall of the cuboid by which the surface waves were meant to be excited.
However, excitation by means of electric field was not successful and eventually the waves were excited mechanically either by ambient vibrational noise or by periodically swinging the whole cryostat in micrometer scale. The second IDC was then used to monitor the static fluid level in the cell.
The wave resonances generated by ambient vibrational noise were found from the Fourier transform of the IDC data processed by an oscilloscope.
Alternatively, when the waves were generated by swinging the cryostat with a series of different frequencies, the IDC signal with 10~ms time resolution was further processed with a lock-in amplifier. For details, see Ref.~\cite{Manninen2013}.

The IDCs were microfabricated elsewhere
by A.~J.~Niskanen, Aalto University, Microfabrication group.
Sapphire substrate ($0.3~\textrm{mm}\times10~\textrm{mm}\times10~\textrm{mm}$) cut orthogonal to the c-axis was patterned with
740 micro strips (spacing 5~$\mu$m, width 5~$\mu$m) and the effective area of the IDC was 7.4~mm wide and 4.6~mm high.
The patterned deposited layer of aluminum was 50~nm thick with additional 15~nm layer of chromium to improve adhesion.
The pattern was superconducting below 1.3~K.
Each IDC was thermalized with a copper mesh (see Fig.~\ref{fig:cell}) 
which was glued (Stycast~1266) to the back of the sapphire plate with an extension bolted to the cell wall.
Superconducting niobium-titanium wires (filament diameter, 62~$\mu$m; bare diameter with copper matrix, 100~$\mu$m) 
were glued to the copper mesh, and 
the superconducting filaments whittled out under the copper matrix were bonded to the contact pads of the IDC with multiple aluminum wires.
The thermalization mesh passed through the narrow channel from the cuboid volume to the annular volume 
making the geometry somewhat ill-defined there.

The experimental setup was originally designed for studies at the melting pressure of helium~\cite{Rysti2014}.
Two quartz tuning fork resonators, usable as thermometers in liquid helium, were thus located at the upper part of the experimental cell, whereas the lower part of the cuboid volume with the IDCs was originally reserved for solid helium.
The helium crystal size could be varied with a separate bellows volume with base temperature not lower than 0.4~mK.
A pressure gauge was also attached to the bellows volume.
The experimental cell and the bellows volume were connected together with a tube (diameter, 2.5~mm; length, 20~cm).
The bellows volume and the quartz tuning forks are not shown in Fig.~\ref{fig:cell} as they were not in use during the experiments with free surfaces.
However, if positioned differently, both the bellows volume and the quartz tuning forks would have been useful assets during the experiments with free surfaces also.

\section{Surface Wave Resonances in Superfluid $^4$He}\label{sec:Data}

We first present here the observed surface wave resonance modes in superfluid $^4$He at the zero temperature limit ($<$~100~mK) as the function of liquid depth in our experimental cell, as driven by the ambient vibrational noise. This is our most complete set of data, which we then analyze in terms of the theoretical models in the following sections. Auxiliary data on the behavior as the function of temperature, the data on superfluid $^3$He, and some observations in the normal fluid He are presented in the latter sections.

\begin{figure}
\includegraphics{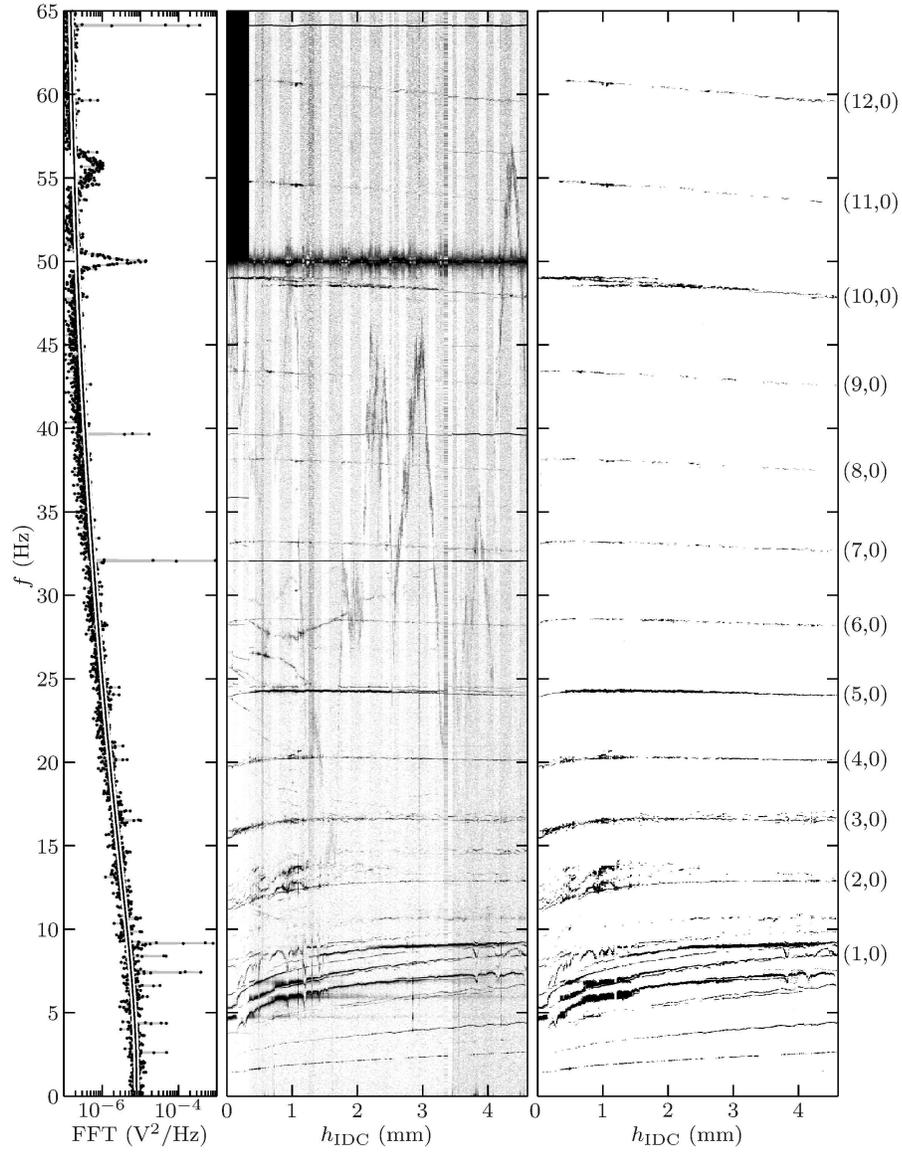}%
\caption{Surface wave resonances in superfluid $^4$He generated by ambient vibrations in the zero temperature limit ($T<100~\mathrm{mK}$). %
Left: A raw FFT spectrum at helium level $h_\mathrm{IDC}=4.4$~mm with fitted background. %
1~V corresponds to 0.17~mm on the IDC. %
Center: Background subtracted FFT spectra with grayscale encoded intensity at various helium levels $h_\mathrm{IDC}$. %
Right: FFT data after filtering and image processing. Only the response due to the liquid surface modes remains visible.
The resonance modes corresponding to the cuboid experimental volume are labeled with indices ($m_x$,$m_y$)
}
\label{fig:Filtered}       
\end{figure}

An example of the Fourier transformed signals is presented in Fig.~\ref{fig:Filtered}. Every such trace is an average of about 20 recordings, each collected for 50~s. The noise bottom had slight dependence on frequency but it remained pretty well constant for the whole duration of the experiment, which took about 24~days. Above the background in the example spectrum, one can observe some sharp peaks below about 10~Hz, at about 32~Hz, 40~Hz, and at about 60~Hz, the 50~Hz contamination at the line frequency and a broader feature slightly above 55~Hz. From an individual sample it is not possible to tell, which of those arise from the oscillations of the helium surface, and which were due to electric disturbances or other artifacts of mechanical origin. It is the continuous filling of the cell and the consistent shift of the relevant resonances as the function of the liquid level, which tells us the difference. We can thus further clean up the data by eliminating any features above the background, that do not depend on the liquid level in the cell. The middle panel in Fig.~\ref{fig:Filtered} shows the background subtracted signals as the function of the liquid depth, and the right panel is a cleaned version, where only the interesting features due to the modes on the liquid surface remain.

Such processing reveals a lot of details not distinguishable at individual spectra. There are about a dozen surface resonances below 15~Hz frequency and a more or less evenly spaced set of 12 resonances up to 60~Hz frequency. No further modes could be observed above that. It is noteworthy that the liquid level dependency of the modes at the highest frequencies was almost entirely due to the slight tilting of the capacitor plates, since the surface dimensions then slightly depended on the liquid level. Were they perfectly aligned, it might have been impossible to distinguish these modes from the background disturbances! The analysis of these sets of resonances is presented after the introduction of the theoretical models in the next section.

The broad feature in the example spectrum in Fig.~\ref{fig:Filtered} at about 55~Hz may deserve a special note. This "bump" was nearly always visible in the data but its position varied seemingly arbitrarily in the course of time from about 20 to 55~Hz. This caused the wandering shadow, that is visible in the middle panel of Fig.~\ref{fig:Filtered}, most clearly at about 3~mm level below 45~Hz and above 4~mm level below 55~Hz. We have no explanation on its origin but it was definitely not due to any action of the superfluid in the experimental cell. The vertical shadows in the middle panel of Fig.~\ref{fig:Filtered} were caused by slight variations of the noise bottom of the spectra. It also appears that the ambient vibration levels at our experimental location were exceptionally high during the measurements at about 1~mm liquid level, as the signals from the motion of the surface appear more intense as usual during that period of time.

\section{Models for Surface Resonances}\label{sec:Models}

\begin{figure}\sidecaption
\includegraphics{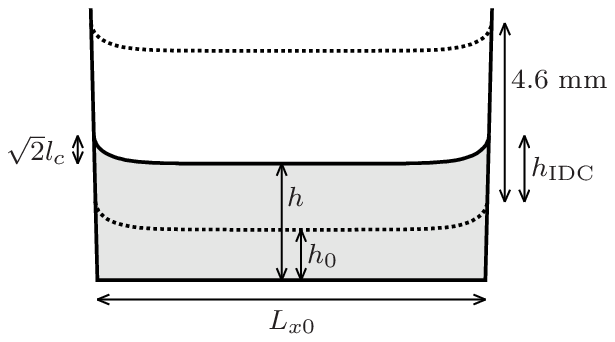}%
\caption{Free surface in the cuboid volume. The lowest measurable helium level is $h=h_0$ when $h_\mathrm{IDC}=0$~mm.
For the highest measurable level $h_\mathrm{IDC}=4.6$~mm.
The capillary raise is $\sqrt{2} l_c=\sqrt{2\gamma/(\rho g)}.$
Volume width at the bottom is $L_{x0}$}
\label{fig:geometry}       
\end{figure}

Standing waves on the free fluid surface are modeled assuming
incompressible, irrotational, homogeneous and inviscid fluid.
The parameters characterizing the geometry of the cuboid volume are shown in Fig.~\ref{fig:geometry}.
In Cartesian coordinates the free surface profile can be written in the form $z_\mathrm{surf}(x,y,t)=h+\eta(x,y)+\xi(x,y,t)$,
where $\eta(x,y)$ describes the stationary meniscus and $\xi(x,y,t)=\zeta(x,y)\tau(t)$ represents a standing wave mode on the surface with $\tau(t)=a\sin(\omega t)$.
For a semi-infinite horizontal fluid surface bound by a wall, that is aligned with the $y$-axis, the shape of the meniscus $\eta(x)$ is obtained from~\cite{Landau1987Fluid}
\begin{eqnarray}
\frac{x-x_0}{2l_c}&=&\frac{1}{2}\mathrm{arccosh}\left(\frac{2 l_c}{\eta}\right)
-\sqrt{1-\left(\frac{\eta}{2 l_c}\right)^2}
\textrm{,}
\label{eq:SurfProfile}
\end{eqnarray}
where $l_c=\sqrt{\gamma/(\rho g)}$ is the capillary length,
and $\gamma$, $\rho$ and $g$ are the surface tension, mass density and gravitational acceleration, respectively.
The parameter $x_0$ is included for setting the proper contact angle with the fluid.
For a vertical wall with zero contact angle $\int \eta\,dx=l_c^2$.

For the surface tension in $^4$He we use $\gamma=375~\mathrm{\upmu J/m^2}$~\cite{Roche97} and thus $l_c=0.513$~mm.
For $^3$He we use $\gamma=155.7~\mathrm{\upmu J/m^2}$~\cite{Iino85a} and $l_c=0.440$~mm.

\subsection{Basic Analytical Model}\label{subsec:ResonanceModels}
Let us begin with a basic model,
where the meniscus effect on the boundaries of the free surface is neglected.
Under that assumption the low amplitude standing surface wave resonances in a rectangular liquid pool occur at frequencies $f=\omega/(2\pi)$ as
\begin{eqnarray}
\omega^2_\mathrm{basic} &=& \left(1+\frac{\gamma}{\rho g}k^2\right)gk\tanh (kh)
\textrm{,}\label{eq:BasicDispersion}\\
k^2&=&\left(\frac{m_x\pi}{L_x}\right)^2+\left(\frac{m_y\pi}{L_y}\right)^2%
\textrm{,}~~~
m_x,m_y\in\mathbb{N}_0%
\textrm{,}
\label{eq:kBasicCuboid}
\end{eqnarray}
where
$h$, $L_x$, and $L_y$ are the liquid depth, and the surface length and width, respectively.
Further, we have the surface wave profile $z_\mathrm{surf}(x,y,t)=h+\xi(x,y,t)$, the scalar potential $\phi(x,y,z,t)$ [defining the velocity field $\mathbf{v}(x,y,z,t)=\nabla\phi$],
and kinetic and potential energies $E_\mathrm{kin}$ and $E_\mathrm{pot}$ as follows:
\begin{eqnarray}
\xi(x,y,t)&=&\cos\!\left(\frac{m_x\pi}{L_x}x\right)\cos\!\left(\frac{m_y\pi }{L_y}y\right) \tau(t)
=\zeta(x,y)\tau(t)%
\textrm{,}\label{eq:BasicProfile}\\
\phi(x,y,z,t)&=&\cos\!\left(\frac{m_x\pi}{L_x}x\right)\cos\!\left(\frac{m_y\pi }{L_y}y\right)\frac{\cosh\!\left(kz\right)}{\sinh\!\left(kh\right)} \frac{1}{k}\dot{\tau}(t)
=\psi(x,y,z)\dot{\tau}(t)%
\textrm{,}\label{eq:BasicScalarpotential}\\
E_\mathrm{kin}&=&\int\!\!\!\!\!\int\!\!\!\!\!\int\frac{1}{2}\rho \mathbf{v}^2\,d\mathbf{r}
=\frac{\dot{\tau}^2}{2}\frac{L_x L_y \rho
}{k \tanh(kh) 2^{D} }
=\frac{\dot{\tau}^2}{2} M_\mathrm{kin}%
\textrm{,}
\label{eq:BasicEkin}\\
E_\mathrm{pot}&=&\int\!\!\!\!\!\int\left\{
\frac{1}{2}\rho g\xi^2
+
\frac{1}{2}\gamma\left[\left(\partial_x \xi\right)^2+\left(\partial_y \xi\right)^2\right]
\right\}\,dx\,dy\nonumber\\
&=&\frac{\tau^2\omega^2}{2}\frac{L_x L_y (\rho g+\gamma k^2)}{\omega^2 2^{D}}
 = \frac{\tau^2\omega^2}{2}M_\mathrm{pot}
\label{eq:BasicEpot}
\textrm{,}
\end{eqnarray}
where the time dependencies are explicitly separated as the factor $\tau(t)=a\sin(\omega t)$.
The parameter $D$ depends on the indexes $m_x$ and $m_y$: having either $m_x=0$ or $m_y=0$ implies $D=1$, and with both $m_x\geq1$ and $m_y\geq1$ one has $D=2$.
From the Euler-Lagrange equation, $\tau$ being the generalized coordinate,
\begin{eqnarray}
0&=&\frac{d}{dt}\left(\frac{\partial L}{\partial \dot{\tau}}\right)-\frac{\partial L}{\partial \tau}
\textrm{,}~~~
L=E_\mathrm{kin}-E_\mathrm{pot}
\label{eq:EulerLagrange}
\end{eqnarray}
it follows that $M=M_\mathrm{kin}=M_\mathrm{pot}$ for the harmonic oscillator -type equations
Eqs.~(\ref{eq:BasicEkin}) and (\ref{eq:BasicEpot}),
in agreement with Eq.~(\ref{eq:BasicDispersion}).

Similar equations can also be written for the annular volume in cylindrical coordinates.
However, it is simpler to describe such a narrow annular volume with the equations derived for the rectangular geometry without any loss of accuracy --- only the boundary condition in one direction becomes periodic due to the solutions wrapping up to themselves around the circumference.

\subsection{Analytical Model with Meniscus Correction}\label{subsec:ResonanceMeniscus}
In order to account for the meniscus effect 
Roche et al.~\cite{Roche97} treated the meniscus $\eta(x,y)$ in a wide well ($m_y=0$, $L_y \gg L_x$) as additional mass on the free surface with the profile $z_\mathrm{surf}(x,y,t)=h+\eta(x,y)+\zeta(x,y)\tau(t)$.
Surface displacement, scalar potential and thus also the total potential energy are assumed to be of the same form as Eqs.~(\ref{eq:BasicProfile}), (\ref{eq:BasicScalarpotential}) and (\ref{eq:BasicEpot}) without the meniscus corrections.
These assumptions are valid only when $kl_c\ll1$:
the meniscus height $\eta\sim l_c$ should be smaller than the depth of the layer where the velocity field is significant, which is given by $k^{-1}$.
However, kinetic energy increases due to the moving additional mass in the meniscus as
\begin{eqnarray}
E_\mathrm{kin}&=&
\frac{\dot{\tau}^2}{2}\frac{L_x L_y \rho
}{2k \tanh(kh) }
\left[1+4k\frac{\gamma}{\rho g}\frac{1}{L_x}\tanh(kh)\right]
\textrm{,}
\label{eq:RocheEkin}
\end{eqnarray}
and the Euler-Lagrange equation Eq.~(\ref{eq:EulerLagrange}) can be used to obtain the resonance frequencies
\begin{eqnarray}
\omega^2 &=& \frac{\omega^2_\mathrm{basic}}
{1+4k\frac{\gamma}{\rho g} \frac{1}{L_x}\tanh (kh)}
\textrm{.}\label{eq:RocheDispersion}
\end{eqnarray}
The original result by Roche et al.~dealt with just one of the two menisci on a straight channel whereas the above equation takes both of the opposite walls into account.

In the same spirit as above it is possible to include also the two additional menisci on the side walls in the case when $m_y=0$ but both $L_x$ and $L_y$ are finite:
\begin{eqnarray}
\omega^2 &=& \frac{\omega^2_\mathrm{basic}}
{1+4k\frac{\gamma}{\rho g}
\left\{
 \frac{1}{L_x}\tanh (kh)
 +\frac{1}{2L_y}
  \left[
   \tanh (kh)+\coth (kh)
  \right]
\right\}
}
\label{eq:RocheFiniteLxLy}
\textrm{.}
\end{eqnarray}
In an annular volume menisci exist on the sides only; thus for the case with periodic boundaries one gets
\begin{eqnarray}
\omega^2 &=& \frac{\omega^2_\mathrm{basic}}
{1+2k\frac{\gamma}{\rho g} 
\frac{1}{L_y}\left[\tanh (kh)+\coth(kh)\right]
}
\label{eq:RocheFiniteLy}
\textrm{.}
\end{eqnarray}

\subsection{Numerical Model with Meniscus}\label{subsec:ResonanceModels}
\begin{figure}\sidecaption
\includegraphics{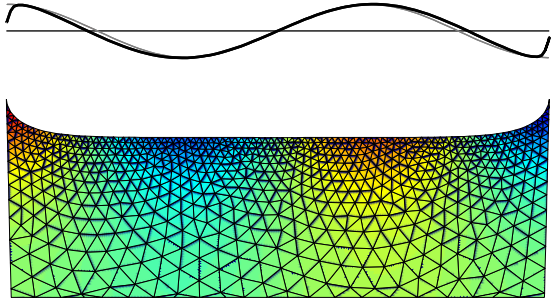}%
\caption{(Color online) (Bottom) scalar potential $\psi(x,z)$ calculated with FEM in 2D according to %
Eqs.~(\ref{eq:FemLaplace}), (\ref{eq:FemWalls}) and (\ref{eq:FemFreeSurf}). %
(Top) the black line illustrates the exaggerated free surface deviation from the equilibrium in the normal direction $\mathbf{\hat{n}}$. %
The sinusoidal form is shown for comparison as the thin gray line. %
The mesh for the model was generated with aid of the code described in Ref.~\cite{MeshFEM2004}}%
\label{fig:FEMexample}
\end{figure}%
The model described in the previous section is limited to wavelengths much longer than the capillary length.
To extend the validity of the analysis to shorter wavelengths and for comparison with cases solvable by other means we have performed numerical computations by the finite-element method (FEM).
It is then straightforward to deal with the geometry altered by the menisci, but it is not so obvious, how to incorporate all effects of the surface tension into the system of equations to be solved. We follow a partly phenomenological approach, where the modified geometry due to the surface tension is built into the model to begin with but the influence of the surface tension to the energy balance is incorporated as a scaling factor for the potential energy in the system. This results in eigenvalues scaled by factors depending on the full solution of the wave profile. This approach was proven valid in cases, that can be compared with analytical solutions.

For the starters the spatial scalar potential $\psi(x,z)$ and the resonance frequency $\omega_0$ are computed in a geometry with curved free surface from the eigenvalue equation
\begin{eqnarray}
\nabla^2\psi&=&0%
\textrm{,}
\label{eq:FemLaplace}\\
{\mathbf{\hat{n}}}\cdot\nabla\psi&=&0
~~~~\textrm{on solid walls,}
\label{eq:FemWalls}\\
{\mathbf{\hat{n}}}\cdot\nabla\psi&=&\frac{\omega_0^2}{g}\frac{1}{\sqrt{1+(\partial_x\eta)^2}}\psi
~~~~\textrm{on free surface,}
\label{eq:FemFreeSurf}
\end{eqnarray}
where ${\mathbf{\hat{n}}}$ is the surface normal vector.
An example solution corresponding to $m_x=3$ is shown in Fig.~\ref{fig:FEMexample}.

\begin{figure}
\includegraphics{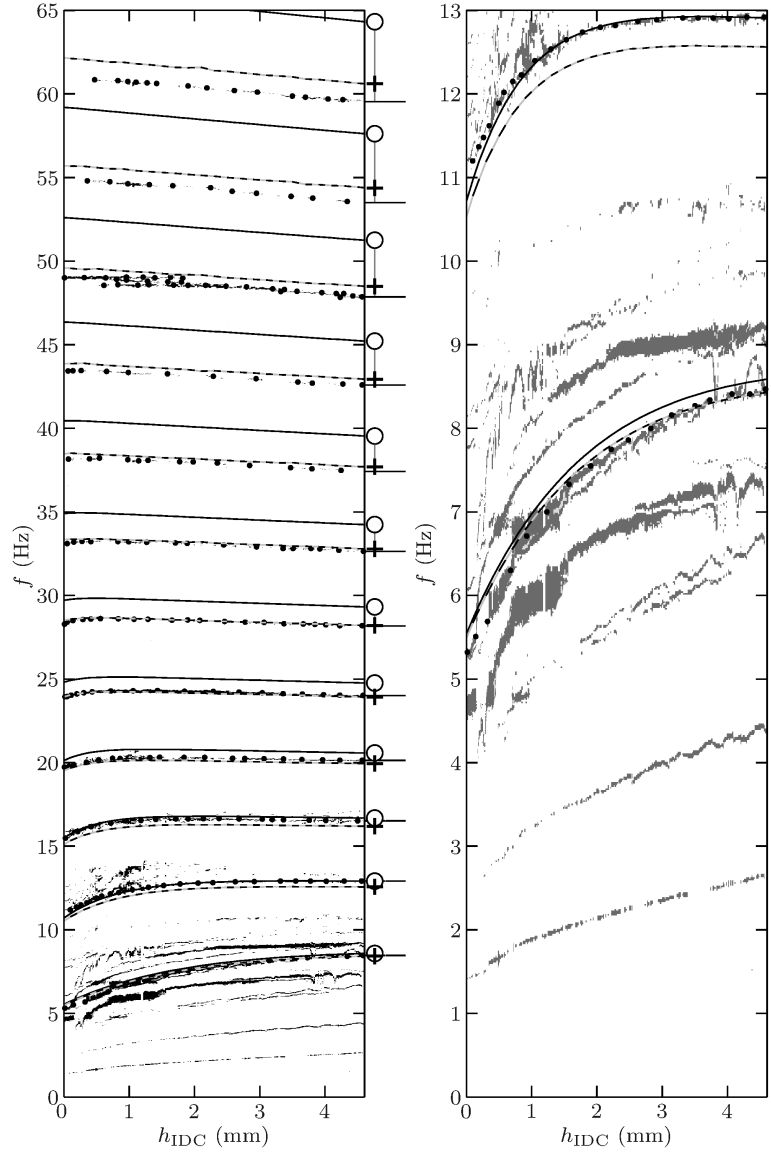}%
\caption{Left: FFT spectra of the surface waves in superfluid $^4$He generated by ambient vibrations %
in the zero temperature limit ($T<100~\mathrm{mK}$) %
at various helium levels $h_\mathrm{IDC}$. %
Solid lines represent eigenfrequencies in the cuboid volume without meniscus correction %
according to Eq.~(\ref{eq:BasicDispersion}). %
Dashed lines are based on 2D FEM calculations according to Eq.~(\ref{eq:FemResFreq}). %
The measured resonance frequencies are highlighted with dots.
Frequency differences between each model and the data are shown as circles and plus-symbols.
Here $\gamma=375~\mathrm{\upmu J/m^2}$, %
$L_{x0}=10.0~\mathrm{mm}$, %
$\theta=2.75^\circ$, %
and $h_0=1.3~\mathrm{mm}$. %
Right: Frequencies up to 13~Hz enlarged}
\label{fig:OPQQcuboid}       
\end{figure}

As the final step to obtain the actual resonance frequencies $\omega$ from the Euler-Lagrange equation Eq.~(\ref{eq:EulerLagrange}), surface tension is included as an extra potential energy term, while assuming that the numerically calculated scalar potential remains unchanged.
The resulting resonance frequencies $\omega$ are attained from
$\psi$ and $\omega_0$ as
\begin{eqnarray}
\omega^2=\omega_0^2\left(
1+\frac{\gamma}{\rho g}
\frac
{\int\!\!\!\int_\mathrm{surf}\frac{(\partial_x \psi)^2}{[1+(\partial_x \eta)^2]^{3/2}}\,dx\,dy}
{\int\!\!\!\int_\mathrm{surf}\psi^2\,\,dx\,dy}
\right)
\textrm{.}
\label{eq:FemResFreq}
\end{eqnarray}
Without the meniscus this modeling principle is exact, as the scalar potential in Eq.~(\ref{eq:BasicScalarpotential}) does not depend on the surface tension at all.
The situation is not quite as clear, when dealing with the curved surfaces, but the anticipated magnitude of the resulting error is not significant.
This model will be discussed in more detail in the
Ph.D.~thesis by Matti~S.\ Manninen.

\section{Comparison of Experiments and Theory}\label{sec:4He}

Let us now further examine the observed surface wave modes in superfluid $^4$He studied below 100~mK while the cell was continuously filled up.

The resonance frequencies corresponding to the geometry of the cuboid volume are presented in Fig.~\ref{fig:OPQQcuboid}.
Altogether twelve resonances can be identified with mode indices from ($m_x=1$,~$m_y=0$) to ($m_x=12$,~$m_y=0$).
Due to the symmetries of our geometry the capacitors were insensitive to the modes with $m_y>0$.
The lowest measurable helium level $h_0=1.3$~mm was treated as a fitting parameter.
Also the angle $\theta=2.75^\circ$ between the two capacitors was determined from the data.

\begin{figure}[b]\sidecaption
\includegraphics{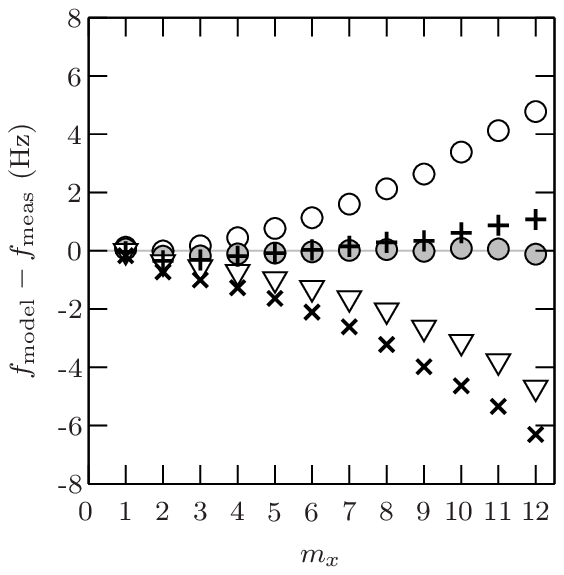}
\caption{Frequency difference between various models $f_\mathrm{model}$ and measured resonance frequencies $f_\mathrm{meas}$
at helium level of $h_\mathrm{IDC}=4.6$~mm.
As in Fig.~\ref{fig:OPQQcuboid}, $\gamma=375~\mathrm{\upmu J/m^2}$, %
$L_{x0}=10.0~\mathrm{mm}$, %
$L_y=10.35~\mathrm{mm}$, and %
$\theta=2.75^\circ$.
Open circles: the basic model without meniscus correction [Eq.~(\ref{eq:BasicDispersion})].
Gray filled circles: the resonance frequencies of the basic model after scaling by factor of $(1+0.014 m_x)^{-1/2}$.
Triangles: Analytical 2D meniscus corrected model [Eq.~(\ref{eq:RocheDispersion})].
Plus-symbols: 2D FEM calculation [Eq.~(\ref{eq:FemResFreq})].
Cross-symbols: 2D FEM calculation [Eq.~(\ref{eq:FemResFreq})] with extra mass meniscus correction on sides
}
\label{fig:Error}       
\end{figure}

Evidently the basic model, Eq.~(\ref{eq:BasicDispersion}), where menisci have been neglected, is not sufficient,
see circles in Figs.~\ref{fig:OPQQcuboid} and~\ref{fig:Error}.
Analytical 2D meniscus correction [Eq.~(\ref{eq:RocheDispersion})] is not any better
as it overestimates the kinetic energy related to the menisci
especially when the condition $k l_c\ll 1$ is not fulfilled
(triangles in Fig.~\ref{fig:Error}).
Moderate agreement with measurements is reached with a numerical 2D model, Eq.~(\ref{eq:FemResFreq}),
where the meniscus effect is taken into account on the two IDC end walls but neglected on the silver sintered side walls
(plus-symbols in Figs.~\ref{fig:OPQQcuboid} and~\ref{fig:Error}).
However, the fluid contact angle on the porous sintered side walls
may also be close to perfectly wetted zero angle and thus the meniscus effect on the side walls should not be omitted.
To illustrate the effect of the menisci on the sides, we have
treated them numerically in $x$-direction [Eq.~(\ref{eq:FemResFreq})]
and in $y$-direction as an extra kinetic energy term, $\frac{1}{2}\rho\int 2 l_c^2 \mathbf{v}^2\,dx$,
see cross-symbols in Fig.~\ref{fig:Error}.
However, this model does not fit the data as the model overestimates the kinetic energy in the menisci.
Evidently, in order to take into account the menisci also on the sides,
a full 3D numerical model should be introduced.

Although the basic model, Eq.~(\ref{eq:BasicDispersion}), does not properly fit the data,
we can estimate the excess energies
from the frequency difference between the simple model and the measurements.
We scale the kinetic energy of the basic model Eq.~(\ref{eq:BasicEkin}) by a factor of $(1+A m_x)$ whereas the potential energy remains as in Eq.~(\ref{eq:BasicEkin}).
As the result, according to the Euler-Lagrange equation Eq.~(\ref{eq:EulerLagrange}),
the resonance frequencies in Eq.~(\ref{eq:BasicDispersion}) are scaled by the factor of $(1+Am_x)^{-1/2}$.
We get $A=0.014$ for this phenomenological fitting parameter, and the corrected resonances are shown as gray filled circles in Fig.~\ref{fig:Error}.
The corrected resonances are in good agreement with the data for all the twelve modes.

To justify the chosen form of the scaling factor $(1+A m_x)$, we note that it is of the same form as Eq.~(\ref{eq:RocheEkin}), triangles in Fig.~\ref{fig:Error}, though $A=4\pi(l_c/L_x)^2=0.033$ in this case for high helium levels.

All our models deal with some liquid in a free volume bound by solid walls.
However, the bottom and the side walls of our cell are coated with a layer of porous sintered silver powder.
The fitted value for $h_0=1.3~\mathrm{mm}$ in Fig.~\ref{fig:OPQQcuboid} suggests that
the distance from the effective region of the IDC to the bottom is
$h_0+\sqrt{2}l_c \approx 2.0~\mathrm{mm}$, see Fig.~\ref{fig:geometry}.
The measured value for the distance from the effective region of the IDC to the sintered bottom is about 1.4~mm and to the impenetrable copper bottom about 2.0~mm.
Thus we conclude that the sintered walls cannot be considered as solid walls for superfluids.
The sintered layer is not a free volume either but rather a region with lower mass density of the fluid.
For the modes with higher frequency or when the helium level is sufficiently high,
the effect caused by the porous sintered layer should be small. Therefore it is not likely that such effects would explain the difficulty of fitting our data with the presented models.

\begin{figure}
\includegraphics{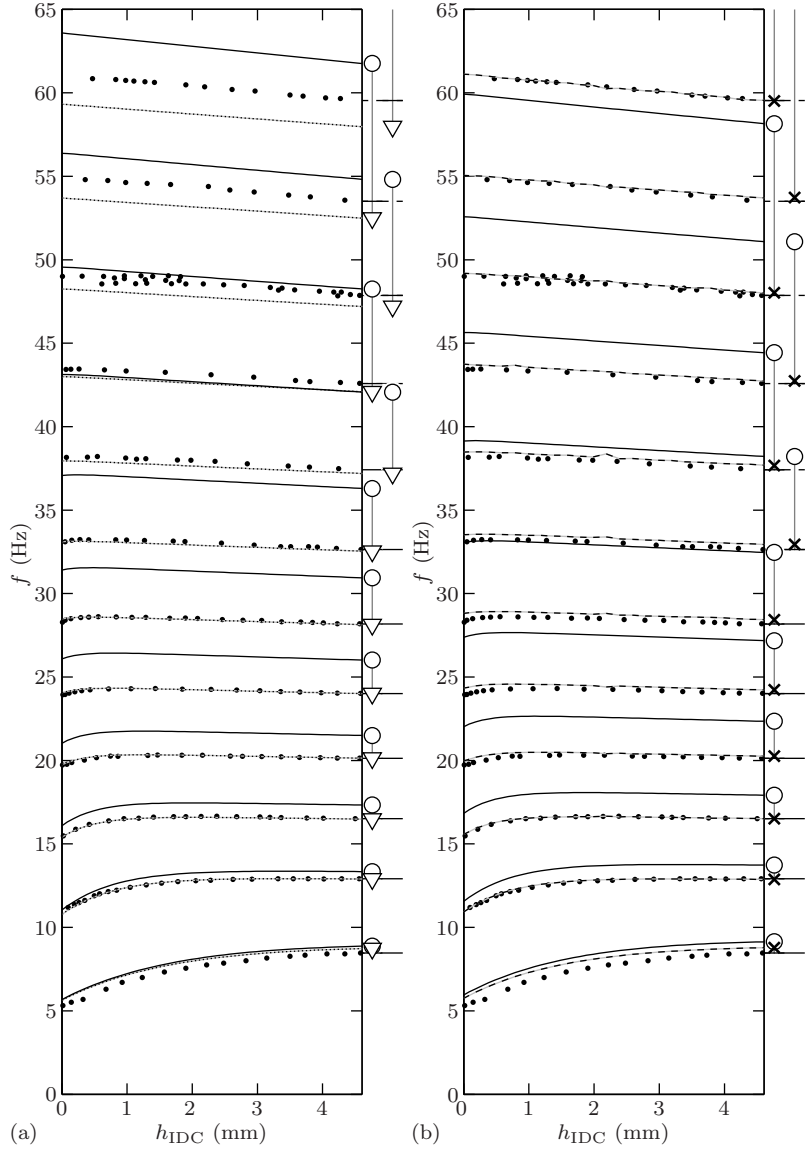}%
\caption{Measured resonance frequencies in superfluid $^4$He (dots).
Solid line and circles are resonance frequencies according to the basic model in Eq.~(\ref{eq:BasicDispersion}).
(a) Dotted lines and triangles represent analytical meniscus corrected 2D model in Eq.~(\ref{eq:RocheDispersion}).
Here $\gamma=375~\mathrm{\upmu J/m^2}$, %
$\theta=2.75^\circ$, %
$h_0=1.22~\mathrm{mm}$, and %
$L_{x0}=9.45~\mathrm{mm}$. %
(b) Dashed lines are based on 2D FEM calculations according to Eq.~(\ref{eq:FemResFreq}) %
with meniscus correction in $y$-direction as additional mass. %
Here $\gamma=375~\mathrm{\upmu J/m^2}$, %
$\theta=2.75^\circ$, %
$h_0=1.22~\mathrm{mm}$, and %
$L_y=10.35~\mathrm{mm}$, %
but $L_{x0}=9.0~\mathrm{mm}$ %
}
\label{fig:cuboidAB}       
\end{figure}

Figs.~\ref{fig:cuboidAB}(a) and~\ref{fig:cuboidAB}(b) are shown to demonstrate that our models can reproduce the measured data if some model parameters are set free.
In Fig.~\ref{fig:cuboidAB}(a) the analytical meniscus corrected 2D model [Eq.~(\ref{eq:RocheDispersion})]
is fitted to the data from $m_x=2$ to $m_x=6$, when $kl_c=(m_x \pi /L_x)l_c<1$, resulting in the fitting parameters $L_{x0}=9.45~\mathrm{mm}$ and $h_0=1.22~\mathrm{mm}$.
Another model, where the meniscus effect was calculated numerically in $x$-direction [Eq.~(\ref{eq:FemResFreq})]
and as an extra kinetic energy term in $y$-direction, is shown in Fig.~\ref{fig:cuboidAB}(b).
Also this model fits well to the data if we set $L_{x0}=9.0~\mathrm{mm}$ though the measured value is about $L_{x0}=10.0~\mathrm{mm}$.

\subsection{Low Frequency Modes and Temperature Dependencies}\label{sec:LF}

The main cuboid volume of our experimental cell should support only one observable primary mode of the free surface below 10~Hz frequency. In contrast, the low frequency part of our data was decorated by several resonances in that region. At least some of those are obviously modes with the largest amplitude in the annular volume surrounding the cuboid volume. We can treat the annular modes according to the argument given in the theory section, although the connection with the cuboid volume through the narrow channels reserved for the wiring etc. cannot be treated in any easy fashion. Such modes apparently produce significant oscillations at the face of the IDCs, though we were not able to perform any simple analysis producing good correspondence with the actual results in this regard.

\begin{figure}
\includegraphics{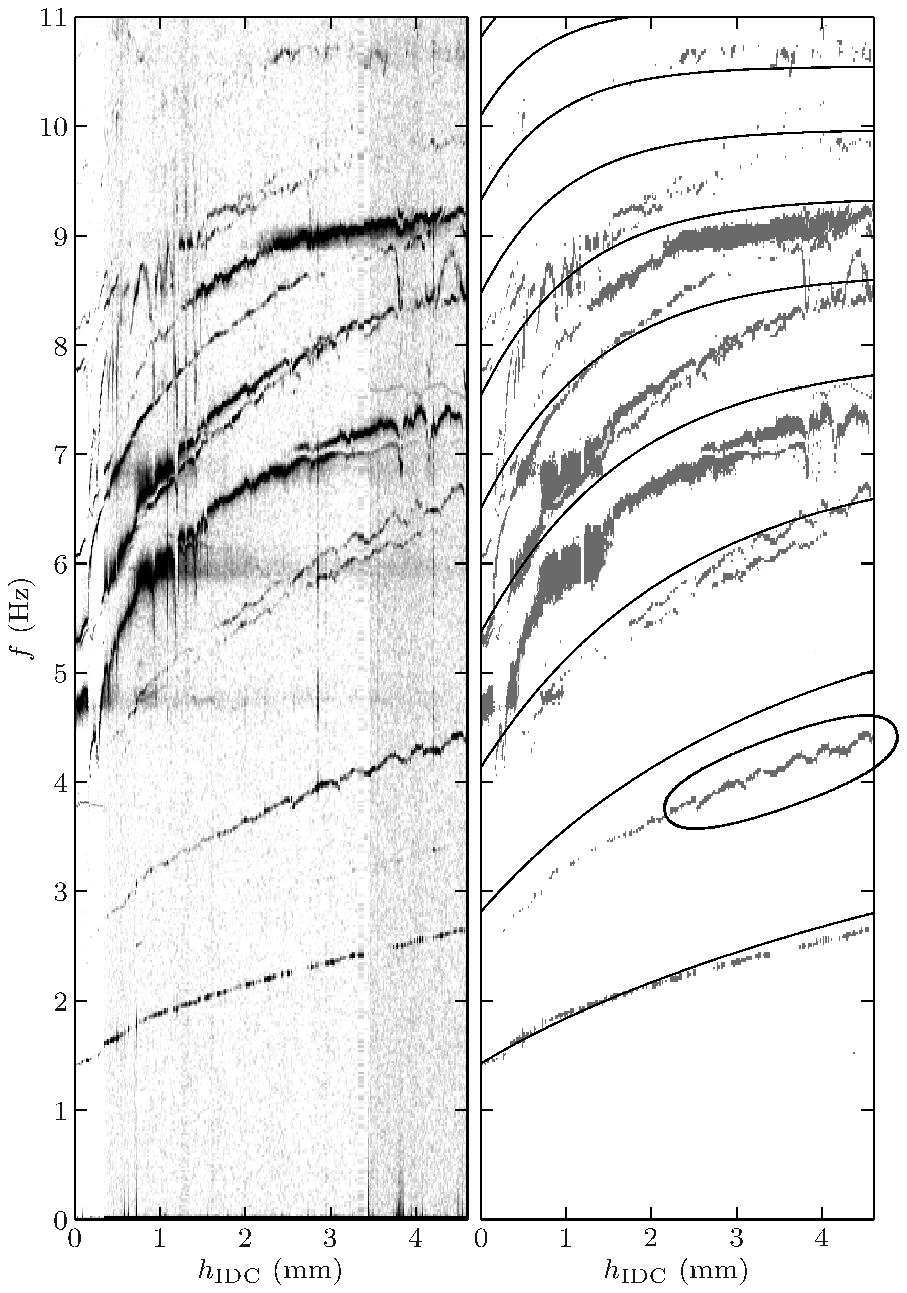}%
\caption{FFT spectra of the surface waves in superfluid $^4$He generated by ambient vibrations in the zero temperature limit ($T<100~\mathrm{mK}$) %
at various helium levels $h_\mathrm{IDC}$. %
The same data before~(left) and after~(right) image processing. %
Solid lines represent meniscus corrected eigenfrequencies in the annular volume %
according to Eq.~(\ref{eq:RocheFiniteLy}) with %
$\gamma=375~\mathrm{\upmu J/m^2}$, %
$L_x=\pi(25.0~\mathrm{mm}+27.6~\mathrm{mm})/2$, %
$L_y=1.3~\mathrm{mm}$, %
and $h_0=1.75~\mathrm{mm}$.
Periodicity of the mode marked with an ellipse is related to geometry
}
\label{fig:OPQQannular}       
\end{figure}

Again, the meniscus effect has to be taken into account.
An analytical meniscus correction on the side walls is given by
Eq.~(\ref{eq:RocheFiniteLy}),
where the condition $k l_c \ll 1$ is better fulfilled
for the long annular channel than for the 1~cm long cuboid volume.
However, in reality the equilibrium free surface profile is more complicated and poorly known
since the menisci are affected by
the thermalization grid of the IDCs
(two strips of copper mesh passing through the annular volume).

As shown in Fig.~\ref{fig:OPQQannular}, eight annular modes can be distinguished,
though the overall correspondence is not as good as for the cuboid volume.
In addition, some modes seem split both in the annular volume
and in the cuboid volume, see Figs.~\ref{fig:OPQQannular} and~\ref{fig:OPQQcuboid}.

We also shortly comment on the artificial periodicity in some of the resonance frequencies, see, as an example, the mode with the second lowest frequency in Fig.~\ref{fig:OPQQannular}.
This 0.38~mm periodicity probably reflects the vertical periodicity of the IDC thermalization mesh fed through the annular volume.

\begin{figure}
\includegraphics{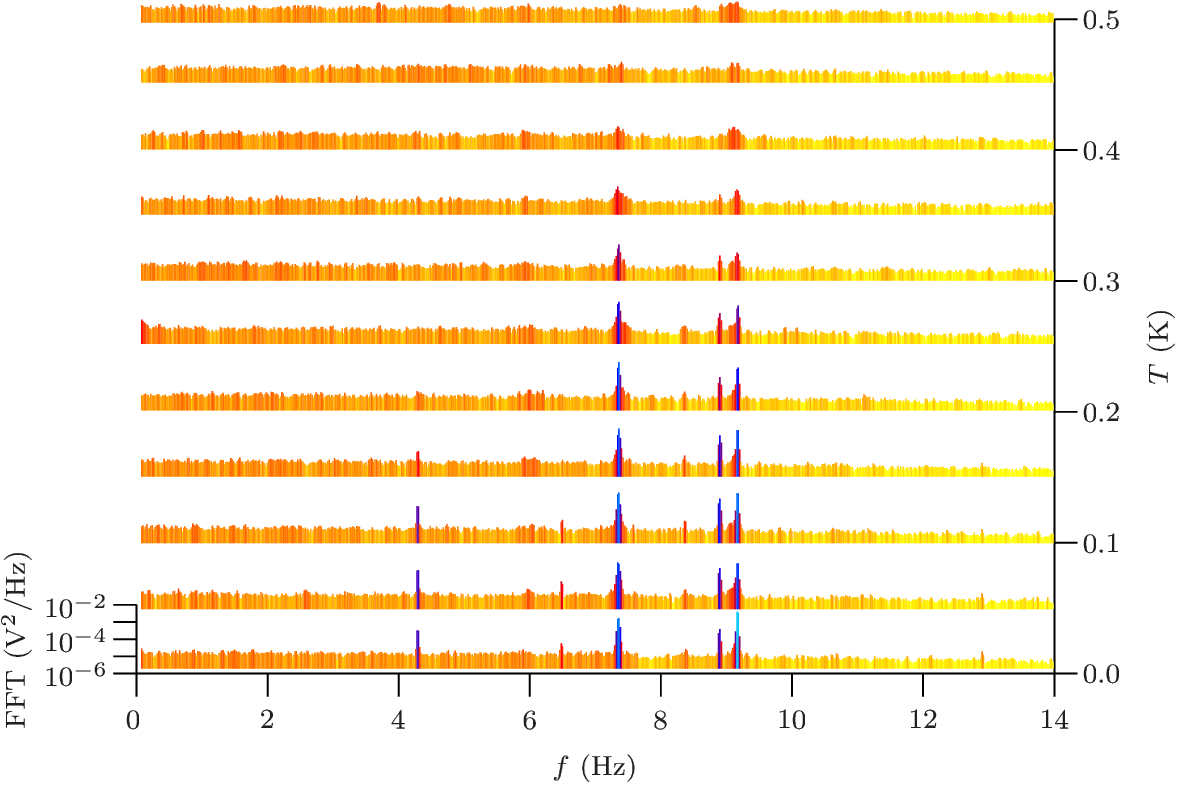}
\caption{(Color online) Frequency spectra of the surface waves in superfluid $^4$He excited by ambient vibrations at various temperatures.
Helium level was $h_\mathrm{IDC}=4.1~\mathrm{mm}$. 
1~V corresponds to 0.17~mm. 
The color code represents the amplitude}
\label{fig:4HeTFFT}       
\end{figure}

\begin{figure}
\includegraphics{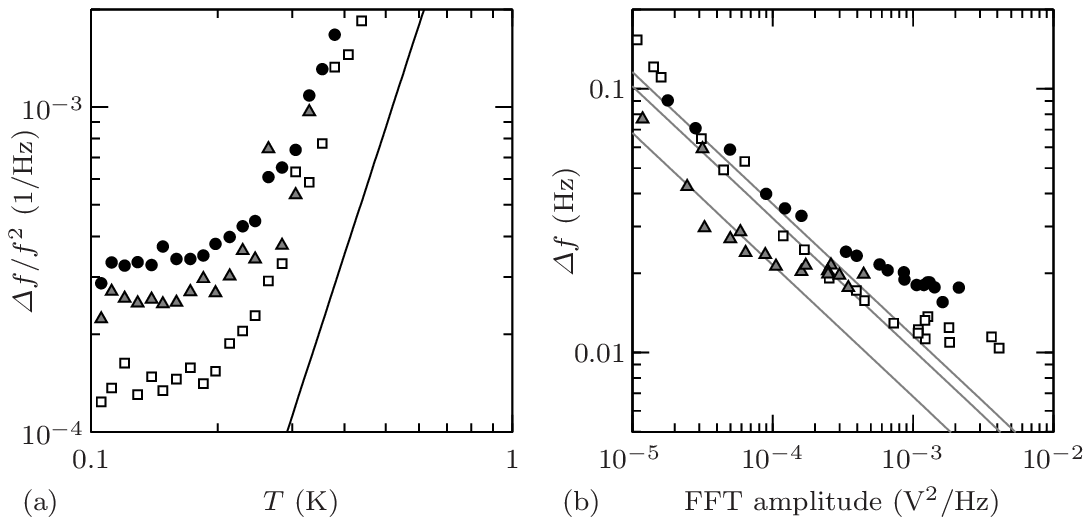}%
\caption{(a) Resonance widths $\Delta f$ of the surface waves excited by ambient vibrations in superfluid $^4$He %
at various temperatures at frequencies of %
$f=7.36~\mathrm{Hz}$ (circles), %
$8.89~\mathrm{Hz}$ (triangles) and %
$9.16~\mathrm{Hz}$ (squares).
Examples of the raw frequency spectra are shown in Fig.~\ref{fig:4HeTFFT}.
Solid line is the expected damping due to thermal quasiparticles.
Helium level was $h_\mathrm{IDC}=4.1~\mathrm{mm}$. %
(b) Amplitude dependency of $\Delta f$ with expected proportionality (gray lines). %
1~V corresponds to 0.17~mm}
\label{fig:4HeT}       
\end{figure}

In these measurements we could not observe any temperature dependencies of resonance frequencies, 
which would have reflected changes in surface tension due to ripplons, quantized surface excitations~\cite{Atkins1953},
see the raw frequency spectra in Fig.~\ref{fig:4HeTFFT}.
Instead, our focus was more on the temperature dependency of resonance widths $\Delta f$.
Temperature dependencies of $\Delta f$ of three resonance modes are shown in Fig.~\ref{fig:4HeT}(a).
As expected, the resonance amplitude $A$ is inversely proportional to the resonance width $\Delta f$ with constant excitation,
see Fig.~\ref{fig:4HeT}(b). 
If we assume this proportionality ($A\Delta f=\mathrm{constant}$), then the integral of the FFT spectrum over the resonance
is inversely proportional to the resonance width, $\int \mathrm{FFT}\,df\propto A^2 \Delta f \propto (\Delta f)^{-1}$.
With the FFT integrals the resonance width data shown in Fig.~\ref{fig:4HeT}(a) can be somewhat extended at
both ends when $\Delta f$ is too narrow or too wide to be determined directly.
Attenuation is discussed in details in another paper~\cite{Manninen14},
where also the theoretical line shown in Fig.~\ref{fig:4HeT}(a) is explained.
However, after revision described 
in the Ph.D.~thesis by Matti~S.\ Manninen,
the theoretical estimate for the damping should have been smaller by factor of two,
and the proper relation for the resonance width is
\begin{equation}
 Q=\frac{f}{\Delta f}=
 \frac{15(\rho g+\gamma k^2) h^3 u^4}{8\pi^6f(k_B T)^4}
 \approx\frac{15\rho g h^3 u^4}{8\pi^6f(k_B T)^4}
 \textrm{,}
\end{equation}
where $u$ is speed of sound in $^4$He.

\section{Experiments on $^3$He}\label{sec:3He}

Experiments on superfluid $^3$He are technically very challenging but bear interest due to the special properties of the superfluid states of $^3$He. Our results here are much less systematic but interesting comparison with the results on $^4$He can still be made.

It is not so easy to give proper readings for the temperature and the zero temperature limit is barely within reach at the zero pressure of our experiments on $^3$He.
We remind, that our cell was originally designed for studies of helium crystals, so that
the installed quartz tuning fork thermometers were useless as they were located above the free surface of helium.
The platinum NMR thermometer was only moderately thermally coupled to the experimental volume and its readings saturated to a value around 0.4 mK.
While the sample was cooled down by lowering the magnetic field $B$ of the copper nuclear cooling stage, the temperature was deduced from the demagnetization field assuming adiabaticity of the process, $T_\mathrm{ns}\propto B$.

The best independent reference point for such low temperatures is the superfluid transition temperature $T_c$ of $^3$He, which could be observed by the interdigital capacitors themselves.
When DC bias voltage was applied over the capacitor, the increase of the capacitance indicates that the higher the DC bias voltage the larger proportion of the surface of the capacitor was covered with helium both below and above $T_{c}$.
This effect was greatly enhanced in superfluid thus clearly showing the onset temperature.
As a curiosity we point out, that even in an almost empty cell
the IDC attracted substantial amount of helium remaining in the porous heat exchangers,
when sufficiently high DC voltage was applied.

\begin{figure}[b]\sidecaption
\includegraphics{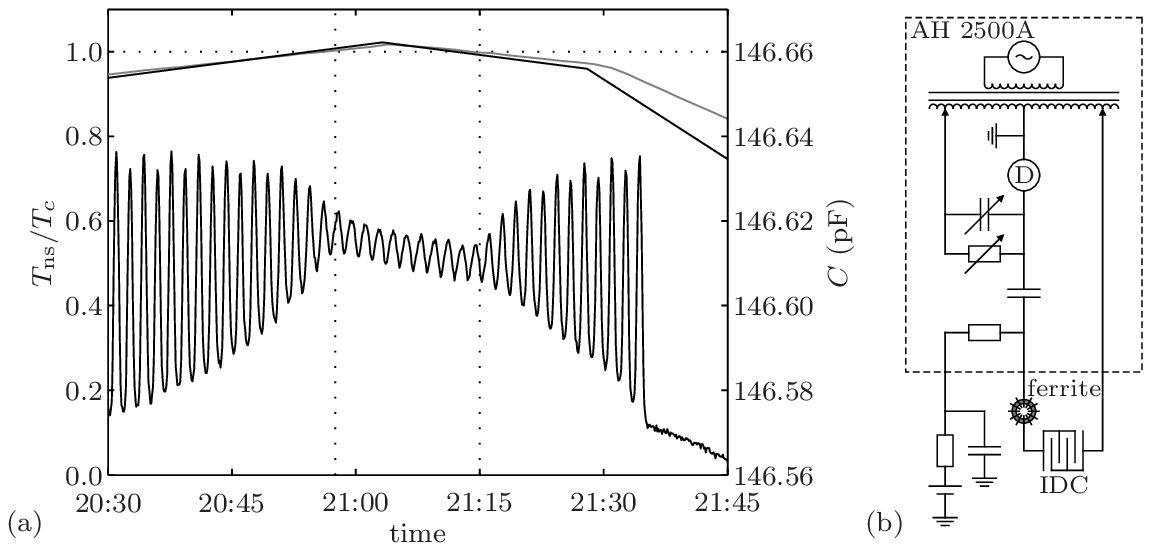}
\caption{(a) Temperature $T_\mathrm{ns}$ deduced either from the platinum nuclear magnetic resonance thermometer (gray line) 
or from the magnetic field of the copper nuclear stage (black solid line). %
Superfluid transition temperature $T_c$ of $^3$He serves as a fixed calibration point (dotted lines), %
which can be seen from the IDC data (oscillatory line) as helium is attracted towards high voltage capacitors %
more effectively below $T_c$. %
(b) Circuit diagram to find $T_c$
}
\label{fig:IDCTc}
\end{figure}

The superfluid transition is demonstrated in Fig.\,\ref{fig:IDCTc}(a), where
DC bias voltage was sinusoidally varied between $-20$~V and $+20$~V at $5$~mHz. 
The measurement circuit with a 1~kHz capacitance bridge is shown in Fig.\,\ref{fig:IDCTc}(b).
The ferrite ring protected the capacitor against the noisy sensing channel of the bridge, which caused significant heating at the lowest temperatures without such filtering.
Our value for the transition temperature $T_c=1.07$~mK is about 10\% higher than often found from literature~\cite{Greywall86}.
The transition temperature was deduced from the demagnetization field which was calibrated with the platinum NMR and other thermometers ($^3$He vapor pressure, calibrated germanium resistor and noise thermometer) at higher temperature.
Also our measured values for the phase transition temperatures $T_c$, $T_\mathrm{AB}$ and $T_N$ at the melting pressure were about 10\% higher
than suggested by PLTS-2000~\cite{PLTS2000}.
Such difference persisted despite of our most careful attempts to perform proper calibrations for our thermometers using the accepted standards, namely the vapor pressure of $^3$He, as reference.

Surface waves in $^3$He were studied with three helium levels only since
continuous filling through the filling line at temperatures far below 1~mK caused excessive heat load to superfluid $^3$He and to the copper nuclear cooling stage.
In order to fill the cell continuously, the compressible bellows volume should have been located below the experimental volume.

\begin{figure}
\includegraphics{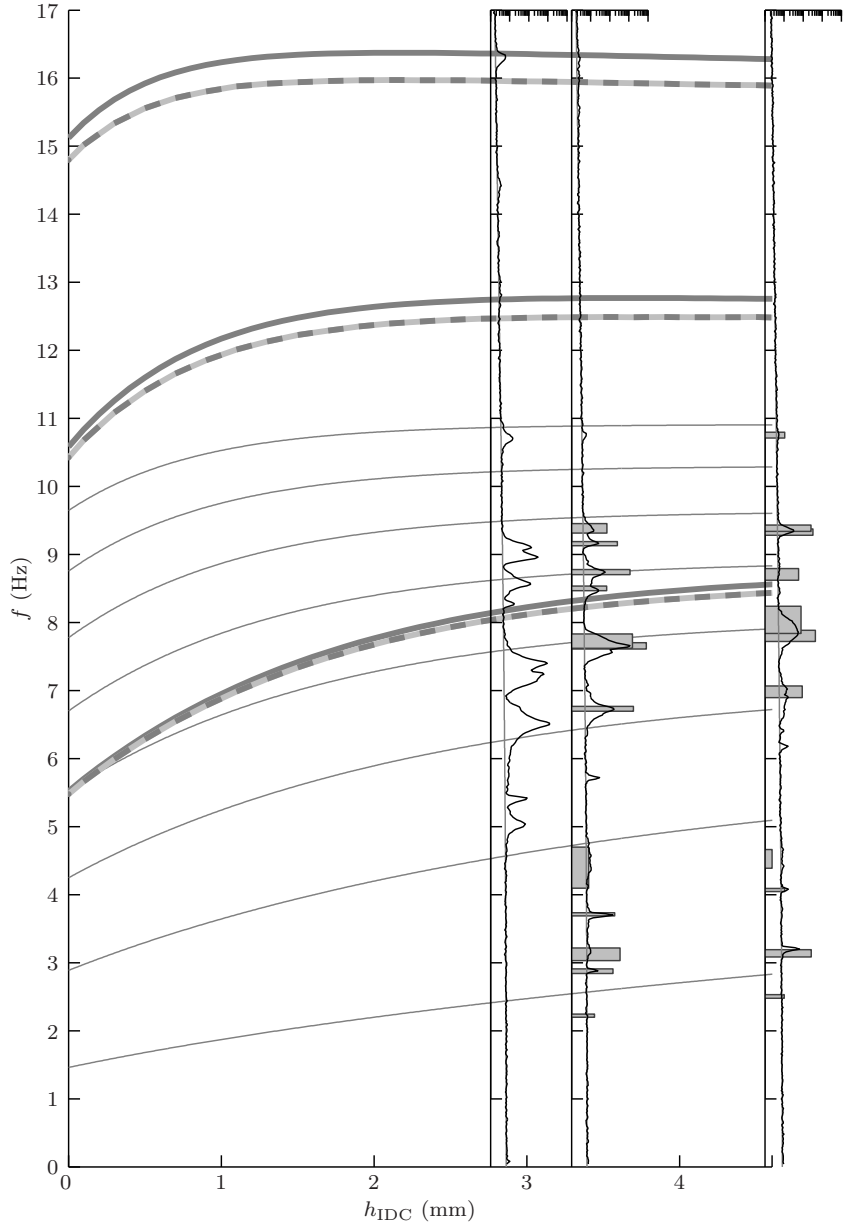}%
\caption{Surface wave measurements in superfluid $^3$He with three helium levels, $h_\mathrm{IDC}=2.8$, $3.3$ and $4.6~\mathrm{mm}$. %
FFT spectra generated by ambient vibrations (black lines). %
Fitted resonance widths and relative amplitudes of the actively excited waves (gray bars),
see an example of the fits to raw data in Fig.~\ref{fig:rockS3He}(a). %
Thick gray lines represent eigenfrequencies in the cuboid volume without meniscus correction %
according to Eq.~(\ref{eq:BasicDispersion}). %
Thick gray dashed lines are based on 2D FEM calculations according to Eq.~(\ref{eq:FemResFreq}) with %
$L_x=10.0~\mathrm{mm}$, %
$\theta=2.75^\circ$, %
and $h_0=1.3~\mathrm{mm}$.
Gray solid lines represent meniscus corrected eigenfrequencies in the annular volume %
according to Eq.~(\ref{eq:RocheFiniteLy}) with %
$L_x=\pi(25.0~\mathrm{mm}+27.6~\mathrm{mm})/2$, %
$L_y=1.3~\mathrm{mm}$, %
and $h_0=1.75~\mathrm{mm}$. %
Here $\gamma=155.7~\mathrm{\upmu J/m^2}$}
\label{fig:3HeOAB}       
\end{figure}

Frequency spectra with the three $^3$He levels are shown in Fig.~\ref{fig:3HeOAB}
together with calculated frequencies corresponding to the modes in
the cuboid volume, Eq.~(\ref{eq:FemResFreq}), and
the annular volume, Eq.~(\ref{eq:RocheFiniteLy}).
The waves were excited either by ambient vibrational noise or by actively swinging the whole cryostat.
As the resonances were measured only with three helium levels in rather complex geometry,
it is difficult to reliably associate the measured and calculated frequencies with each other.
The most distinct resonance at 16~Hz corresponds to $m_x=3$ in the cuboid volume though it was
clearly seen only with the lowest helium level.

\begin{figure}
\includegraphics{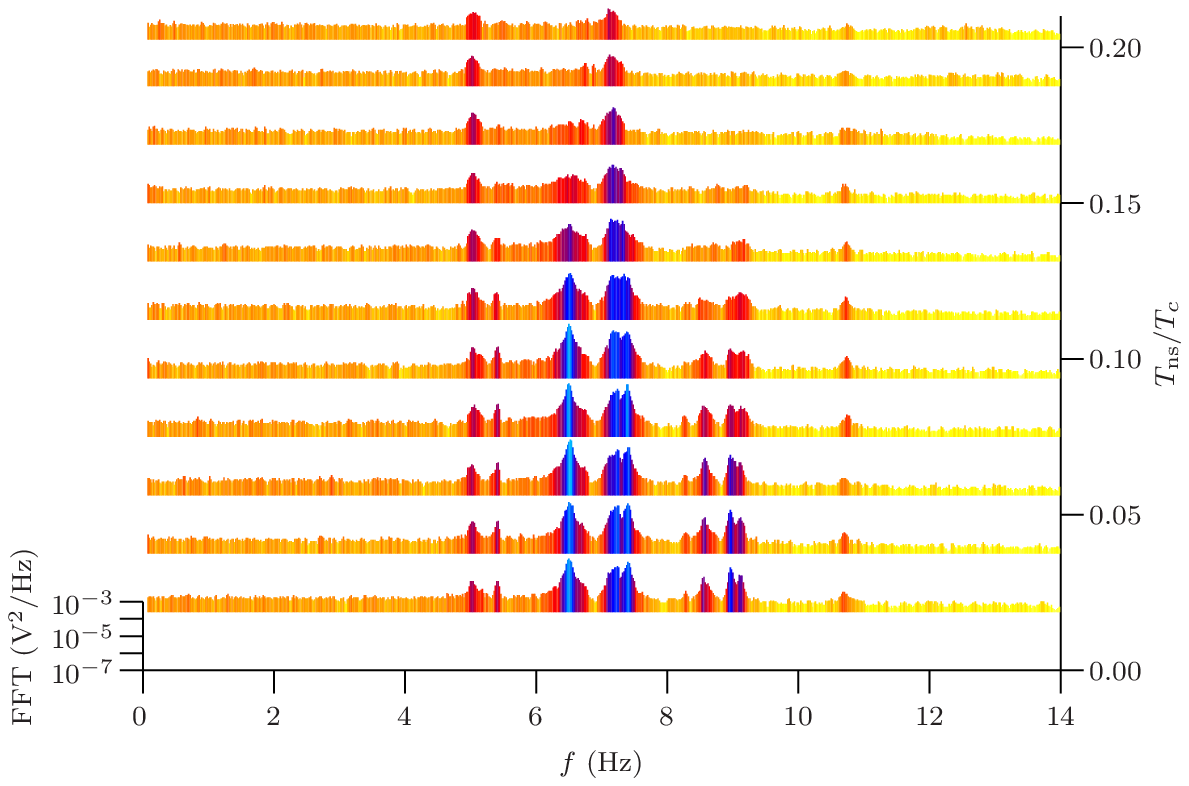}
\caption{(Color online) Frequency spectra of the free surface waves in superfluid $^3$He at various temperatures $T_\mathrm{ns}$ of the nuclear stage.
Here the waves were excited mechanically by noisy environment.
Helium level was $h_\mathrm{IDC}=2.8~\mathrm{mm}$.
1~V corresponds to 0.2~mm.
The amplitude is highlighted with the color code}
\label{fig:3HeTFFT}       
\end{figure}

As shown in Fig.~\ref{fig:3HeTFFT} the surface waves in superfluid $^3$He excited by ambient vibrations of the environment
vanished at temperatures $T>T_c/4$ being observable again only above $T\sim50$~mK.  
The temperature was deduced from the demagnetization field as the quartz tuning fork thermometers were located above the free surface. With full cell the lowest temperatures measured with the tuning forks were about $0.12T_c$ at the melting pressure~\cite{Todoshchenko2014}.
However, during the experiments on the free helium surface, heat leak to the experimental cell was reduced
as the tube connecting the experimental cell with the bellows volume was empty. 

\begin{figure}
\includegraphics{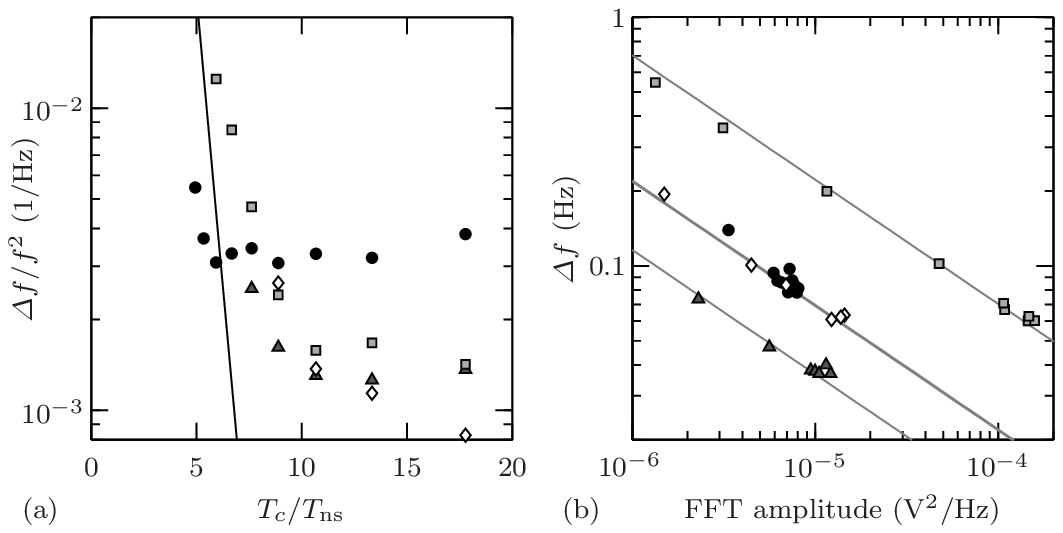}%
\caption{(a) Resonance widths $\Delta f$ of the surface waves excited by ambient vibrations in superfluid $^3$He %
at various temperatures $T_\mathrm{ns}$ at frequencies of %
$f=5.04~\mathrm{Hz}$ (circles), %
$5.41~\mathrm{Hz}$ (triangles), %
$6.50~\mathrm{Hz}$ (squares) and
$8.57~\mathrm{Hz}$ (diamonds).
The raw spectra are shown in Fig.~\ref{fig:3HeTFFT}.
Helium level was $h_\mathrm{IDC}=2.8~\mathrm{mm}$. %
(b) Amplitude dependency of $\Delta f$ with expected proportionality (gray lines).
1~V corresponds to 0.2~mm}
\label{fig:3HeT}       
\end{figure}

The resonance widths in superfluid $^3$He at various temperatures are shown in Fig.~\ref{fig:3HeT}(a)
together with the expected quasiparticle damping~\cite{Manninen14}
\begin{equation}
 Q=\frac{f}{\Delta f}=\frac{(\rho g+\gamma k^2)h^3\exp[E_\Delta /(k_B T)]}{8\pi^2 f p_F^4}
 \approx\frac{\rho g h^3\exp[E_\Delta /(k_B T)]}{8\pi^2 f p_F^4}
\mathrm{,}
\end{equation}
where $E_\Delta$ and $p_F$ are the energy gap and Fermi momentum, respectively.
The expected inverse proportionality between $\Delta f$ and amplitude is confirmed in Fig.~\ref{fig:3HeT}(b).
In contrast to $^4$He, $\Delta f$ was wider than our frequency resolution even at the lowest temperatures.
Thus there was no need to determine $\Delta f$ indirectly by integrating the FFT spectrum over the resonance.

\begin{figure}
\includegraphics{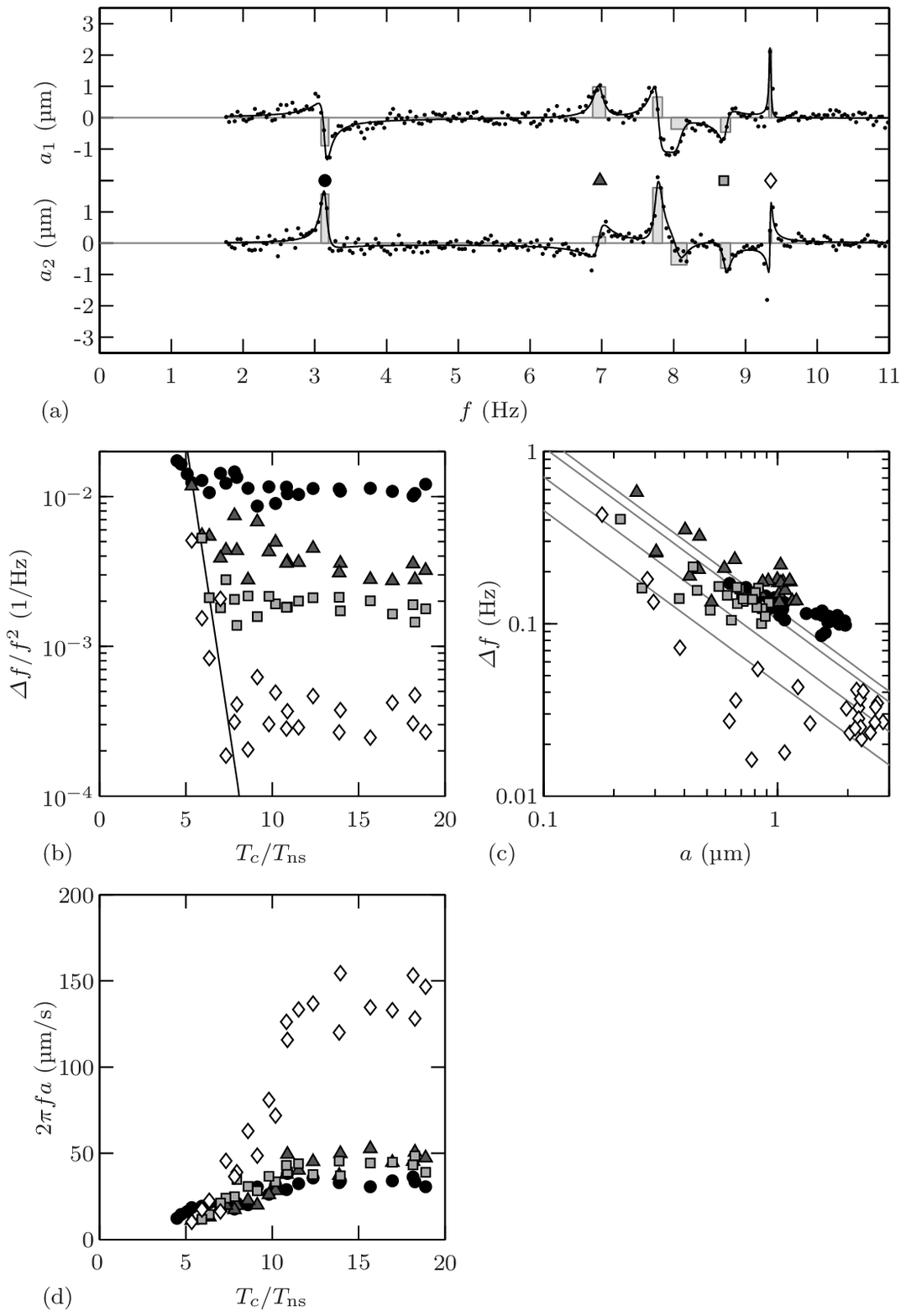}%
\caption{(a) Phase-sensitive detection of surface waves in superfluid $^3$He %
actively generated by swinging the cryostat. Detection frequency $f$ was five times higher than the excitation frequency. %
Black solid lines are fits to the measured data (black dots). %
Resonance widths and amplitudes of the fitted curves are visualized as gray bars. %
Here $h_\mathrm{IDC}=4.6~\mathrm{mm}$ and the temperature of the nuclear stage was in the zero temperature limit for the superfluid, $T_\mathrm{ns}=0.03T_c$. %
(b) Temperature dependency of resonance width $\Delta f$. %
Solid line is the expected damping due to thermal quasiparticles. %
(c) Amplitude dependency of $\Delta f$.
(d) Temperature dependency of the wave amplitude $a$ at a resonance. %
The amplitude $a$ is multiplied by resonance frequency $f$ which corresponds to vertical surface velocity on the IDC}
\label{fig:rockS3He}       
\end{figure}

An alternative technique, phase sensitive active mechanical drive, can be used to study the temperature dependence of $\Delta f$,
see Fig.~\ref{fig:rockS3He}.
The waves were excited with non-sinusoidal mechanical swinging
and measured with higher harmonic frequencies.
This method is very slow, since each individual frequency must be measured separately, see Fig.~\ref{fig:rockS3He}(a).
Temperature dependence of $\Delta f$ is weak [Fig.~\ref{fig:rockS3He}(b)],
though the amplitude is temperature dependent even below $0.1T_c$ [Fig.~\ref{fig:rockS3He}(d)].
This can be seen also in Fig.~\ref{fig:rockS3He}(c) where $\Delta f$ saturates from the expected inverse
proportionality to the amplitude at the lowest temperatures.
In addition, the frequency widths of the modes with different frequencies saturate at different temperatures
to value about 0.1~Hz for most of the modes.
In principle, the heat load caused by the mechanical excitation could depend on frequency,
though with full cell the superfluid $^3$He temperature measured directly with quartz tuning fork thermometers
was not excitation frequency dependent.
In Fig.~\ref{fig:rockS3He}(d) the wave amplitude is scaled by frequency,
which corresponds to the vertical velocity of the wave.
Even at the lowest temperatures the wave velocity is very low, $v p_F\ll k_B T$,
which was assumed in Ref~\cite{Manninen14} when calculating the damping due to quasiparticles shown in Fig.~\ref{fig:rockS3He}(b).

\begin{figure}
\includegraphics{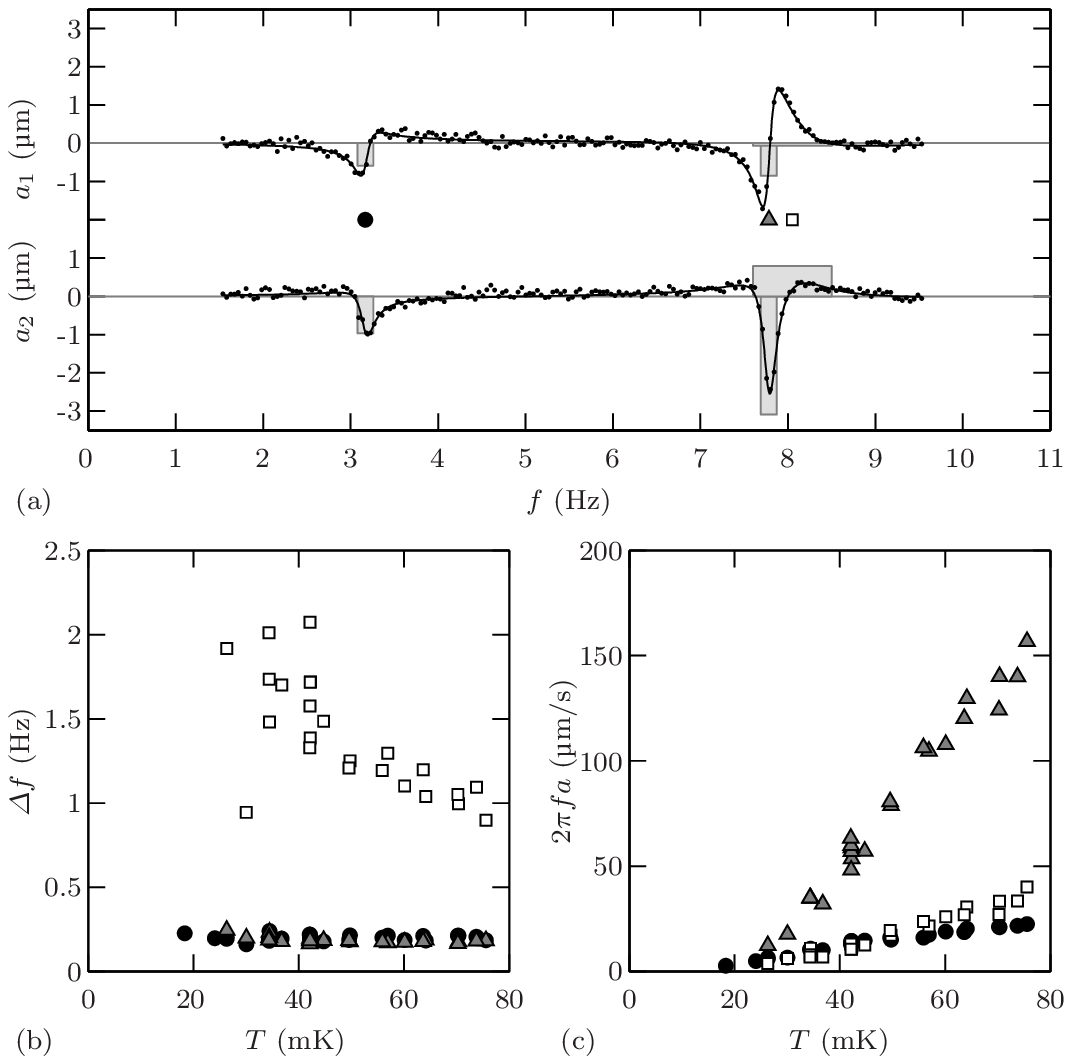}%
\caption{(a) Phase-sensitive detection of surface waves in normal $^3$He %
actively generated by swinging the cryostat. 
Detection frequency $f$ was three times higher than the excitation frequency.
Black solid lines are fits to the measured data (black dots). %
Resonance widths and amplitudes of the fitted curves are visualized as gray bars. %
Here $T=75~\mathrm{mK}$ and $h_\mathrm{IDC}=3.1~\mathrm{mm}$. %
(b) Temperature dependency of resonance width $\Delta f$. %
(c) Vertical RMS speed of the resonance on the IDC deduced from the RMS amplitude $a$ multiplied by resonance frequency~$f$}
\label{fig:rockN3He}       
\end{figure}

For comparison with the data measured in superfluid $^3$He, Fig.~\ref{fig:rockN3He} illustrates the measurements in $^3$He above the superfluid transition temperature. Here again the waves were excited by swinging the whole cryostat pneumatically.
The temperature dependence is seen merely in amplitude [Fig.~\ref{fig:rockN3He}(c)]
while $\Delta f$ is either temperature independent or the dependence is weak [Fig.\ \ref{fig:rockN3He}(b)].
This is not what was expected. As for the waves excited by ambient vibrational noise,
the prediction was, that $\Delta f$ and amplitude would have had inverse correspondence.

\section{Conclusions}\label{sec:Conclusions}

Surface waves were studied both in superfluid $^4$He and $^3$He.
Superfluid $^4$He level was raised continuously below 100~mK and 12 frequency modes up to 60~Hz were observed corresponding to the geometry of the main cuboid experimental volume.
In addition, there were several modes originating from the surrounding annular volume coupled to the cuboid volume via two narrow channels.
In superfluid $^3$He the highest resonance frequency was 16~Hz
corresponding to the third mode in the cuboid volume.
The resonances in the annular volume could not be identified
as the waves were studied only with three distinct $^3$He levels.
Moreover, the characteristics of the frequency spectra in $^4$He and in $^3$He differed significantly though
they were supposed to be similar oscillations in the same geometry.

The surface wave resonance frequencies were measured with great precision especially in superfluid $^4$He.
In principle surface tension could have been deduced from those frequencies.
However, this is not straightforward since, besides our nontrivial cell geometry,
the effect of surface tension is twofold.
In addition to the direct surface potential energy
it has an additional effect to the surface shape by forming menisci near the walls.
The measurements were in reasonable agreement with a 2D numerical model where both effects were taken into account.
Also the basic analytical model fits to the data when kinetic energy is scaled by factor of $(1+0.014m_x)$ to effectively account for the meniscus correction.
Better agreement 
is expected with a full 3D numerical model where surface tension and porosity of the sintered layers are properly taken into account.
This may also result a more precise value for the surface tension as the discrepancy among previous measurements is about 6\%~\cite{Roche97,Iino1986,Nakanishi1998,Vicente2002}.

The temperature dependence of the resonance width was studied with two different methods.
When the waves were generated by ambient vibrational noise, the resonance width was inversely proportional to amplitude as expected.
However, for the waves in $^3$He actively generated by swinging the whole cryostat
the temperature dependency of the width was weak even though
the amplitude was temperature dependent.
Above $T_c$ the width was essentially temperature independent.

Admittedly, our experimental arrangement was not optimal for producing easily interpretable data on free surface resonances in superfluids, as it was initially designed for studies of 
melting-freezing waves, or crystallization waves, on the superfluid-solid interface.
Since the experiments performed at very low temperatures are usually not easily reconfigured, any modifications to the present system were not undertaken. Further experiments using a more suited setup designed particularly for this purpose would probably clarify some of our unfit results. The presented study was performed, nevertheless, since it was a straightforward extension to the solid phase experiment and since very few or no similar studies exist.
Moreover, the basic features are similar for the waves both on the free surface and on the mobile superfluid-solid interface.
Further discussion on both phenomena can be found from the Ph.D.\ thesis by Matti S.\ Manninen.


%
%

\begin{acknowledgements}
We thank J.-P. Kaikkonen, A.~J.~Niskanen, A.~Ya.~Parshin, V.~Peri, A.~Ranni, A.~Salmela, A.~Sebedash,
and V.~Tsepelin for valuable discussions and assistance.
This work was supported in part by the European Union
FP7/2007–2013, Grant No. 228464 Microkelvin and by the
Academy of Finland, CoE 2012–2017, Grant No. 250280 LTQ.
This research made use of the Aalto University Low Temperature
Laboratory infrastructure. We also acknowledge the grants
from Jenny and Antti Wihuri Foundation and the National
Doctoral Programme in Materials Physics.
\end{acknowledgements}

\bibliographystyle{spphys}       
\bibliography{Name}   

\begin{thebibliography}{10}
\providecommand{\url}[1]{{#1}}
\providecommand{\urlprefix}{URL }
\expandafter\ifx\csname urlstyle\endcsname\relax
  \providecommand{\doi}[1]{DOI \discretionary{}{}{}#1}\else
  \providecommand{\doi}{DOI \discretionary{}{}{}\begingroup
  \urlstyle{rm}\Url}\fi

\bibitem{Manninen14}
M.S. Manninen, J.~Rysti, I.~Todoshchenko, J.~Tuoriniemi, Phys.~Rev.~B
  \textbf{90}, 224502 (2014)

\bibitem{Eltsov13arxiv}
V.B. {Eltsov}, P.J. {Heikkinen}, V.V. {Zavjalov}, arXiv:1302.0764

\bibitem{Yao00YKI}
W.~Yao, T.~Knuuttila, K.~Nummila, J.~Martikainen, A.~Oja, O.~Lounasmaa,
  J.~Low~Temp.~Phys. \textbf{120}, 121 (2000)

\bibitem{Manninen2013}
M.S. Manninen, J.P. Kaikkonen, V.~Peri, J.~Rysti, I.~Todoshchenko,
  J.~Tuoriniemi, J.~Low~Temp.~Phys. \textbf{175}, 56 (2014)

\bibitem{Rysti2014}
J.~Rysti, M.~Manninen, J.~Tuoriniemi, J.~Low Temp.~Phys. \textbf{175}, 739
  (2014)

\bibitem{Landau1987Fluid}
L.D. Landau, E.M. Lifshitz, \emph{Fluid Mechanics}, 2nd edn. (Pergamon Press,
  1987)

\bibitem{Roche97}
P.~Roche, G.~Deville, N.~Appleyard, F.~Williams, J.~Low Temp.~Phys.
  \textbf{106}, 565 (1997)

\bibitem{Iino85a}
M.~Iino, M.~Suzuki, A.~Ikushima, Y.~Okuda, J.~Low~Temp.~Phys. \textbf{59}, 291
  (1985)

\bibitem{MeshFEM2004}
P.~Persson, G.~Strang, SIAM Review \textbf{46}, 329 (2004)

\bibitem{Atkins1953}
K.R. Atkins, Can.~J.~Phys. \textbf{31}, 1165 (1953)

\bibitem{Greywall86}
D.S. Greywall, Phys.~Rev.~B \textbf{33}, 7520 (1986)

\bibitem{PLTS2000}
R.~Rusby, M.~Durieux, A.~Reesink, R.~Hudson, G.~Schuster, M.~Kühne, W.~Fogle,
  R.~Soulen, E.~Adams, J.~Low~Temp.~Phys. \textbf{126}, 633 (2002)

\bibitem{Todoshchenko2014}
I.~Todoshchenko, J.P. Kaikkonen, R.~Blaauwgeers, P.J. Hakonen, A.~Savin, Rev.
  Sci. Instrum. \textbf{85}(8), 085106 (2014)

\bibitem{Iino1986}
M.~Iino, M.~Suzuki, A.~Ikushima, J.\ Low Temp.\ Phys. \textbf{63}, 495 (1986)

\bibitem{Nakanishi1998}
K.~Nakanishi, M.~Suzuki, J.\ Low Temp.\ Phys. \textbf{113}, 585 (1998)

\bibitem{Vicente2002}
C.~Vicente, W.~Yao, H.J. Maris, G.M. Seidel, Phys. Rev. B \textbf{66}, 214504
  (2002)

\end{thebibliography}

%
%

\end{document}